\documentstyle[vsolj01,graphicx,natbib]{article}

\RequirePackage[T1]{fontenc}

\def\cite{\citealt}

\begin{document}

\title{MP Gem: VY Scl-type star, finally}

\author{Taichi Kato$^1$}
\author{$^1$ Department of Astronomy, Kyoto University,
       Sakyo-ku, Kyoto 606-8502, Japan}
\email{tkato@kusastro.kyoto-u.ac.jp}

\begin{abstract}
MP Gem has long been suspected as a long-period eclipsing
binary, which had not been seen in faint state for
71 years since the discovery in 1944.
Its nature has been a mystery.
Using Public Data Release of Zwicky Transient Facility
observations, I finally reached a conclusion that this
object is a VY Scl-type cataclysmic variable by the detection
of the deep faint state and the rising phase in 2018.
\end{abstract}

   MP Gem was discovered as a variable star
by \citet{hof63an287169}.
\citet{hof63an287169} noted that the object was invisible
on a total of nine plates taken on 1944 February 24/25 and
1944 February 25/26.  \citet{hof63an287169} classified
the variable as ``Algol?'' (eclipsing binary).
No other ``eclipse'' was recorded on Sonneberg plates
according to \citet{ges73VSmpgem}.  \citet{boh12mpgem}
studied further Sonneberg plates between 1981 and 1994
and found no further event.  \citet{boh12mpgem} suggested
that the object could be an eclipsing binary with
a very long period.

   \citet{boh14mpgem} described further details of
the research given in \citet{boh12mpgem} and also provided
time-resolved photometry with a CCD camera together with
observations by Timo Kantolo.  Although short-term variations
were recorded, they could not find a period nor obtain
a hint for the classification of this object.

   \citet{boh17mpgem} continued observations and found
a significant drop in brightness in 2016 September,
first time in 72 years.
The object faded to 19.5 mag on 2016 October 20.
B\"ohme called attention to this object through
the American Association of Variable Star Observers (AAVSO).
At that time, \citet{boh17mpgem} found that the fading
was neutral in color and suspected that the object
would be an R CrB star.  B\"ohme also reported this
finding to VSNET \citep{VSNET} as vsnet-alert 20211
on2016 October 5.\footnote{
http://ooruri.kusastro.kyoto-u.ac.jp/mailarchive/vsnet-alert/20211
}
No further call for attention was issued through VSNET
and I was not aware that this could be an event of
a cataclysmic variable (CV).

   In 2018, I picked up candidates of CVs from
the General Catalogue of Variable Stars
(GCVS, \cite{GCVS51}) using the newly available
Gaia parallaxes and colors (currently \cite{GaiaEDR3}).
MP Gem was one of them and I obtained a light curve
using the All-Sky Automated Survey for Supernovae (ASAS-SN,
\cite{ASASSN}, \cite{koc17ASASSNLC}).
Due to the blending with a nearby star (see \cite{boh17mpgem}),
the deep fading starting from 2016 September was recorded
only as a systematic brightness decrease by $\sim$1.0 mag.
I was not confident about the nature of this object
at that time although the blue Gaia color ($BP$=16.05 and
$RP$=15.88) and the faint absolute magnitude
($M_V \sim +$4.0) were clearly indicative of a CV.

\begin{figure*}
  \begin{center}
    \includegraphics[width=16cm]{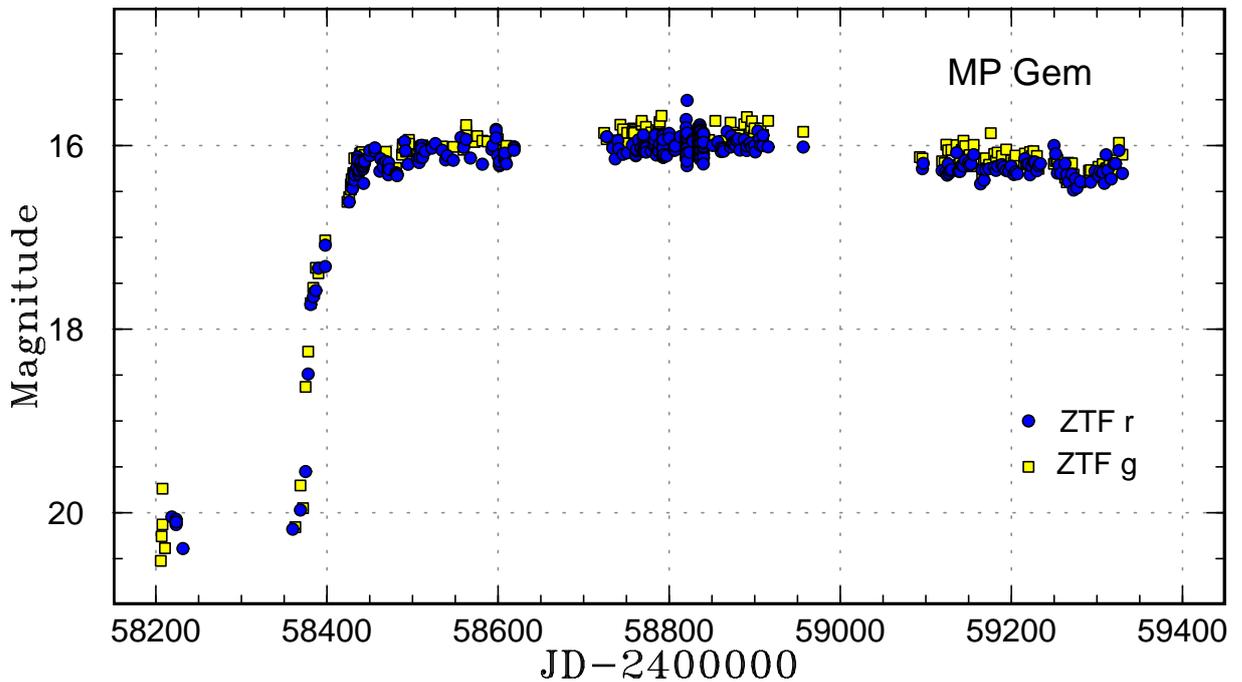}
  \end{center}
  \caption{ZTF light curve of MP Gem.  The light curve and
    blue color are characteristic to a VY Scl-type object.}
  \label{fig:mpgem}
\end{figure*}

   Using Public Data Release 6 of
the Zwicky Transient Facility \citep{ZTF}
observations\footnote{
   The ZTF data can be obtained from IRSA
$<$https://irsa.ipac.caltech.edu/Missions/ztf.html$>$
using the interface
$<$https://irsa.ipac.caltech.edu/docs/program\_interface/ztf\_api.html$>$
or using a wrapper of the above IRSA API
$<$https://github.com/MickaelRigault/ztfquery$>$.},
I found that MP Gem was in deep faint state
in 2018 around 20.0 mag and slowly returned to
the bright state between 2018 August and November
(figure \ref{fig:mpgem}).
This light curve is characteristic to VY Scl-type
novalike CVs, which show fading episodes caused by temporary
decreases in the mass-transfer rate (see \cite{hel01book},
Chapter 12).  MP Gem is now finally classified as
a VY Scl-type object with rather infrequent fading
episodes.

   I noticed that the AAVSO collected a considerable
record of the 2016--2018 event in
the AAVSO International Database\footnote{
   $<$http://www.aavso.org/data-download$>$.
}.  According to their observations, there was a temporary
rise to 17.5 mag in 2017 December to 2018 January.
Such temporary brightening during the faint state
is frequently seen in VY Scl-type objects
(see e.g. \cite{hon04vyscl}).

   Historically, eclipsing binaries, R CrB stars and
VY Scl-type CVs have often been confused.
For an example of V504 Cen, see \citet{kat03v504cen}.
They can usually be easily distinguished by colors
(R CrB stars also show infrared excesses) and
absolute magnitudes when Gaia parallaxes are meaningfully
measured.

\section*{Acknowledgments}

The author is grateful to Naoto Kojiguchi for supplying
a wrapper code for obtaining the ZTF data.

This work was supported by JSPS KAKENHI Grant Number 21K03616.

Based on observations obtained with the Samuel Oschin 48-inch
Telescope at the Palomar Observatory as part of
the Zwicky Transient Facility project. ZTF is supported by
the National Science Foundation under Grant No. AST-1440341
and a collaboration including Caltech, IPAC, 
the Weizmann Institute for Science, the Oskar Klein Center
at Stockholm University, the University of Maryland,
the University of Washington, Deutsches Elektronen-Synchrotron
and Humboldt University, Los Alamos National Laboratories, 
the TANGO Consortium of Taiwan, the University of 
Wisconsin at Milwaukee, and Lawrence Berkeley National Laboratories.
Operations are conducted by COO, IPAC, and UW.

The ztfquery code was funded by the European Research Council
(ERC) under the European Union's Horizon 2020 research and 
innovation programme (grant agreement n$^{\circ}$759194
-- USNAC, PI: Rigault).

\newcommand{\noop}[1]{}\newcommand{\hyphalt}{-}


\begin{thebibliography}{}

\bibitem[{B{\"o}hme}(2012)]{boh12mpgem}
  {B{\"o}hme}, D. (2012) Is {MP Geminorum} an eclipsing binary with a very long
  period?. {\em J.\ American\ Assoc.\ Variable\ Star\ Obs.\,} {\bf 40}, 973

\bibitem[{B{\"o}hme}(2014)]{boh14mpgem}
  {B{\"o}hme}, D. (2014) {Ist} {MP Gem} ein {Bedeckungsver\"anderlicher}?. {\em
  BAV\ Rundbrief\,} {\bf 63}, 146
  (https://www.bav{\hyphalt}astro.eu/rb/rb2014{\hyphalt}3/146.pdf)

\bibitem[{B{\"o}hme}(2017)]{boh17mpgem}
  {B{\"o}hme}, D. (2017) {MP Geminorum} nach 72 {Jahren} wieder im {Minimum}.
  {\em BAV\ Rundbrief\,} {\bf 66}, 14
  (https://www.bav{\hyphalt}astro.eu/rb/rb2017{\hyphalt}1/14.pdf)

\bibitem[{Gaia Collaboration} et~al.(2021)]{GaiaEDR3}
  {Gaia Collaboration} {et~al.} (2021) {Gaia Early Data Release} 3. {Summary}
  of the contents and survey properties. {\em A\&A\,} {\bf 649}, A1

\bibitem[{Gessner}(1973)]{ges73VSmpgem}
  {Gessner}, H. (1973) {Die} veraederlichen {Sterne} der nordlichen
  {Milchstrasse}. {Teil} {XIV}. {\em Ver{\"{o}}ff.\ Sternw.\ Sonneberg\,} {\bf
  7}, 521

\bibitem[Hellier(2001)]{hel01book}
  Hellier, C. (2001) Cataclysmic Variable Stars: How and why they vary (Berlin:
  Springer)

\bibitem[{Hoffmeister}(1963)]{hof63an287169}
  {Hoffmeister}, C. (1963) Mitteilungen {\"u}ber neuentdeckte
  {ver\"{a}nderliche} sterne. {\em Astron.\ Nachr.\,} {\bf 287}, 169

\bibitem[{Honeycutt} and {Kafka}(2004)]{hon04vyscl}
  {Honeycutt}, R.~K., \& {Kafka}, S. (2004) Characteristics of
  high-state/low-state transitions in {VY Sculptoris} stars. {\em AJ\,} {\bf
  128}, 1279

\bibitem[Kato and Stubbings(2003)]{kat03v504cen}
  Kato, T., \& Stubbings, R. (2003) {VY Scl}-type star {V504 Cen}. {\em IBVS\,}
  {\bf 5426}

\bibitem[Kato et~al.(2004)]{VSNET}
  Kato, T., Uemura, M., Ishioka, R., Nogami, D., Kunjaya, C., Baba, H., \&
  Yamaoka, H. (2004) {Variable Star Network}: World center for transient object
  astronomy and variable stars. {\em PASJ\,} {\bf 56}, S1

\bibitem[{Kochanek} et~al.(2017)]{koc17ASASSNLC}
  {Kochanek}, C.~S. {et~al.} (2017) {The All-Sky Automated Survey for
  Supernovae} ({ASAS-SN}) light curve server v1.0. {\em PASP\,} {\bf 129},
  104502

\bibitem[{Masci} et~al.(2019)]{ZTF}
  {Masci}, F.-J. {et~al.} (2019) {The Zwicky Transient Facility: Data
  Processing, Products, and Archive}. {\em PASP\,} {\bf 131}, 018003

\bibitem[{Samus'} et~al.(2017)]{GCVS51}
  {Samus'}, N.~N., {Kazarovets}, E.~V., {Durlevich}, O.~V., {Kireeva}, N.~N.,
  \& {Pastukhova}, E.~N. (2017) General catalogue of variable stars: {Version
  GCVS} 5.1. {\em Astron.\ Rep.\,} {\bf 61}, 80

\bibitem[{Shappee} et~al.(2014)]{ASASSN}
  {Shappee}, B.~J. {et~al.} (2014) The man behind the curtain: {X}-rays drive
  the {UV} through {NIR} variability in the 2013 {AGN} outburst in {NGC 2617}.
  {\em ApJ\,} {\bf 788}, 48

\end{thebibliography}
\end{document}